# Spin correlations in the extended kagome system YBaCo$_3$FeO$_7$

Martin Valldor,[a] Raphaël P. Hermann,[b,c] Joachim Wuttke,[d] Michaela Zamponi,[d] Werner Schweika[b]

[a]*II. Physics Department, University of Cologne, 50937 Cologne, Germany*
[b]*Jülich Center for Neutron Science JCNS and Peter Grünberg Institute PGI, JARA-FIT, Forschungszentrum Jülich, 52425 Jülich, Germany*
[c]*Faculty of Science, University of Liege, 4000 Liege, Belgium*
[d]*Jülich Centre for Neutron Science JCNS, Forschungszentrum Jülich GmbH, Outstation at FRM II, Lichtenbergstraße 1, 85747 Garching, Germany*

Abstract

The transition metal based oxide YBaCo$_3$FeO$_7$ is structurally related to the mineral *Swedenborgite* SbNaBe$_4$O$_7$, a polar non-centrosymmetric crystal system. The magnetic Co$_3$Fe sublattice consists of a tetrahedral network containing kagome-like layers with trigonal interlayer sites. This geometry causes frustration effects for magnetic ordering, which were investigated by magnetization measurements, Mössbauer spectroscopy, polarized neutron diffraction, and neutron spectroscopy. Magnetization measurement and neutron diffraction do not show long range ordering even at low temperature (1 K) although a strong antiferromagnetic coupling (~2000 K) is deduced from the magnetic susceptibility. Below 590 K, we observe two features, a spontaneous weak anisotropic magnetization hysteresis along the polar crystallographic axis and a hyperfine field on the Fe kagome sites, whereas the Fe spins on the interlayer sites remain idle. Below ~50 K, the onset of a hyperfine field shows the development of moments static on the Mössbauer time scale also for the Fe interlayer sites. Simultaneously, an increase of spin correlations is found by polarized neutron diffraction. The relaxation part of the dynamic response has been further investigated by high-resolution neutron spectroscopy, which reveals that the spin correlations start to freeze in below ~50 K. Monte Carlo simulations show that the neutron scattering results at lower temperatures are compatible with a recent proposal that the particular geometric frustration in the *Swedenborgite* structure promotes quasi one dimensional partial order.

Corresponding author:
Martin Valldor
II. Physics Department, University of Cologne
Zülpicher Str. 77
50937 Cologne, Germany
E-mail: valldor@ph2.uni-koeln.de
Fax: 0049-221-470-6708

## I. INTRODUCTION

Geometric frustration, a common phenomenon in many crystal structures, has an important effect in magnetic systems, reducing the stability of ordered antiferromagnetic states and being the origin of various examples of exotic spin physics. One of the most prominent examples of geometric frustration is the "spin-ice" property of pyrochlores ($Dy_2Ti_2O_7$), where the existence of topological magnetic monopoles has been suggested.[1-3] The typical structural features of triangle and tetrahedral networks can be found not only in pyrochlores but in many crystal structures, for example, in garnets, spinels, and in a number of kagome systems, such as the *Herbert-smithite* $ZnCu_3(OH)_6Cl_2$,[4] *Volborthite* $Cu_3V_2O_7(OH)_2 \bullet 2H_2O$,[5] *Magnetoplumbite* $SrCr_{8-x}Ga_{4+x}O_{19}$,[6] and *Jarosite* $(D_3O)Al_3(SO_4)_2(OD)_6$.[7] Magnetic *Swedenborgites*, structurally related to the mineral $SbNaBe_4O_7$,[8-10] also belong to the class of geometrically frustrated magnets containing kagome planes in an alternating stacking with trigonal interlayer sites. According to a recent study, the neutron diffraction patterns of Co and Fe based *Swedenborgites* reveal a broad spectrum of spin freezing phenomena, ranging from long- to short-range magnetic order *i.e.* Néel states to apparent spin-liquid or spin glass states.[11] Magnetic studies[12-14] of the compound $YBaCo_4O_7$ yielded partly controversial results with respect to the transition temperature and the nature of magnetic ordering. According to Chapon *et al.* magnetic long-range order occurs below 110 K,[12] whereas according to Soda *et al.*[13] and Valldor *et al.*[14] only broad magnetic peaks are observed in neutron diffraction data due to short-range ordering of magnetic moments. Recently, Manuel *et al.* measured the diffuse scattering from a $YBaCo_4O_7$ single crystal at 130K,[15] *i.e.* above the ordering temperature. By comparing the observations with Monte Carlo simulations the authors argue that the system is suitably described by an isotropic nearest neighbor Heisenberg model. Interestingly, the isotropic nearest neighbor exchange model for the *Swedenborgite* structure results in spin-correlations with strong one dimensional character, a finding that has been confirmed in a more recent detailed study.[16] The closely related compound $Y_{0.5}Ca_{0.5}BaCo_4O_7$, obtained by partial substitution of the cation Y by Ca, does not show any long-range ordering down to temperatures as low as 1.2 K[17] although the Curie-Weiss constant is huge (~2000 K).[18] The diffuse magnetic scattering of $Y_{0.5}Ca_{0.5}BaCo_4O_7$ closely resembles the scattering from a well-studied model of frustration, *i.e.* the nearest neighbor kagome antiferromagnet, and compares favorably well with the modeled spin correlations of decoupled two dimensional kagome lattices. Apparently the interlayer coupling must be significantly weaker in $Y_{0.5}Ca_{0.5}BaCo_4O_7$ than in $YBaCo_4O_7$. Whereas the absence of long-range order in 2D Heisenberg antiferromagnets is in accordance with the Mermin-Wagner theorem,[19] surprisingly, a complete suppression of long-range order in 3D coupled spin systems was also found in recent model calculations.[16] Apparently, in the case of 3D coupling the disordered ground state has to be attributed to unusually strong geometric frustration effects, which are different from site- or bond-disorder, the typical ingredients for spin glass formation.[20]

An alternate model including 3D spin correlations has been proposed[21] that provides preliminary insights into the spin dynamics. Two components are observed in neutron spectroscopy. One of these corresponds to slow spin relaxations and indicates a freezing of spin configurations below approximately 50 K. The dynamical response related to the second component is temperature independent and corresponds to fast spin relaxations. However, to date, neither structural investigations nor the first results on the spin dynamics in the *Swedenborgites* have delivered a sufficiently consistent understanding.

The substitution of the magnetic ions, for example Co by Fe, provides a further degree of freedom. According to powder neutron diffraction the compound YBaCo$_3$FeO$_7$ does not show an ordered magnetic ground state.[11] This new system is of particular interest because large and phase-pure single crystals can be readily grown, as will be reported below.

An extensive structural description of YBaCo$_3$FeO$_7$ was first reported elsewhere[22] and only a brief description is given here in order to support the understanding of the results below. The atomic ordering in the title compound can be described in two distinct ways, starting either from the (i) anions or the (ii) cations. (i) BaO$_7$ densely packs with the *ABAC* layer ordering, leaving Ba twelve coordinated. Embedded in the oxygen sublattice, the octahedral sites are occupied by Y and tetrahedral sites are filled with Co or Fe (Fig.1a). Three quarters of the Co,Fe-sites constitute *kagome* layers separated by the remaining interlayer sites. (ii) Considering only the cations, it is interesting to see that the metal sites are closely related to the Laves phases (*AB$_2$*), where YBa occupy the *A* and Co$_3$Fe the *B* sites. Hence, a better description of the *Swedenborgite* structure is the close packing of cations (Fig.1b), according to Laves,[23] where the voids are filled with oxygen anions.

In this paper, a suite of experimental methods comprising Mössbauer spectroscopy, X-ray- and neutron scattering on polycrystalline samples, and magnetization measurements on single crystals have been applied to investigate the *Swedenborgite* YBaCo$_3$FeO$_7$, providing new insights into unusual and complex spin correlations, the observation of weak spontaneous spin anisotropy of the kagome spins, and a spin-glass like freezing induced by the onset of spin correlations involving the interlayer sites at low temperatures.

## II. SAMPLE CHARACTERIZATION AND EXPERIMENTAL METHODS

Polycrystalline samples were made from a stoichiometric mixture of Y$_2$O$_3$ (Alfa Aesar 99.99%), BaCO$_3$ (Strem Chem. 99.9%), Co$_3$O$_4$ (Alfa Aesar 99.7%), Fe$_2$O$_3$ (Sigma-Aldrich +99%). After mixing the constituents in an agate mortar, the powder was heated twice at 1100°C in air in a corundum crucible with an intermediate grinding. A second sample was synthesized using $^{57}$Fe$_2$O$_3$ (>95% isotope pure). A Pt crucible was used as reaction vessel for the synthesis of this isotope enriched powder but the heating procedure was the same. Large single crystals were synthesized in a floating-zone mirror-image furnace (FZ-T-10000-H-VI-VP). The polycrystalline seed and feeding rods were heated over night in a corundum crucible in air at 1100 °C, in order to ensure mechanical stability. A steady flow of pure Ar (5 cm$^3$/min) and growth rates between 2 and 10 mm/h were used during the crystal growth process.

The oxygen content of the resulting crystal was investigated indirectly by analyzing the oxidation states of the transition metal ions by x-ray absorption spectroscopy: only Co$^{2+}$ and Fe$^{3+}$ were found,[24] confirming the stoichiometric oxygen content of O$_{7.00(3)}$, as the pristine oxidation states Y$^{3+}$ and Ba$^{2+}$ are expected to be present in the crystal due to their high stability.

Powder X-ray diffraction was performed with a STOE STADI_P in Bragg-Brentano geometry (θ-2θ) using Cu K$\alpha_{1,2}$ radiation (λ = 1.54056, 1.54439 Å) and a position sensitive detector.

All diffraction intensities in the data from the ground single crystal sample can be indexed with the previously reported hexagonal cell.[22] Due to the small size of the sample for the Mössbauer spectroscopy measurement and the uncertain reactivity of $^{57}$Fe$_2$O$_3$ (≈60 mg), small amounts of impurities are present in the powder sample (Fig.2), which could be explained by a few percent of unreacted Y$_2$O$_3$ and the layered perovskite YBa(Co,Fe)$_2$O$_5$. It should be noted that the single crystal sample ($a$ = 6.32261(3), $c$ = 10.2843(1) Å) seems to have slightly larger unit cell parameters, as the diffraction intensities are shifted to lower angles relative to the isotope sample ($a$ = 6.3165(4), $c$ = 10.2749(7) Å); the reason for this is not clear. Hence, the magnetization measurements and Mössbauer spectroscopy presented below were performed on both samples. The relative Bragg intensity differences are most probably due to a preferred orientation of the crystals in the powdered single crystalline sample.

Magnetic investigations were carried out using a vibrating sample magnetometer inside a physical property measurement system (VSM-PPMS, Quantum Design) with fields of up to 1 T in the temperature range 2-1000 K with two different sample holders for either high (300-1000 K) or low temperatures (2-350 K). Zero-field cooling data was obtained by cooling the sample from the highest temperature to the lowest, before applying a measuring field. Accordingly, in field cooling, the sample was exposed to the measuring field during lowering the temperature. All measurements were done on heating. The magnetic relaxation measurement followed a field cooling with 0.1 T from 650 down to 570 K. Subsequently the field was shut off and the magnetization in the sample was monitored as function of time.

Mössbauer spectra were recorded on a constant acceleration spectrometer using a 50 mCi $^{57}$Co@Rh source calibrated with α-iron at 298 K. Spectra below room temperature were obtained with the sample in a helium flow cryostat having a temperature accuracy better than 1%. Spectra between 295 and 480 K were recorded from a 12 mg/cm$^2$ sample mixed with BN in a furnace, with a temperature accuracy of ~5 K calibrated to the α-iron isomer shift. In order to assess the influence of thickness broadening, a second sample with 60 mg/cm$^2$ of natural isotopic abundance was prepared and spectra were recorded at 4.2 and 85 K.

Neutron diffraction with polarization analysis was performed at the DNS instrument[25] of the Jülich Centre for Neutron Science JCNS, outstation FRM II, Munich, Germany. Polarized neutrons with a wavelength of 4.75 Å have been used with subsequent XYZ-polarization analysis[26] in order to separate the magnetic scattering from the structural diffraction. A 30 g powdered single crystal sample was studied between 1.2 and 300 K.

The spin dynamics and relaxation has been studied by neutron spectroscopy using the backscattering instruments BASIS (SNS, Oak Ridge National Laboratory)[27] and SPHERES (FRMII, Munich, Germany).[28] BASIS is a time-of-flight backscattering spectrometer, where a dynamic range of +/−100 µeV has been studied with an energy resolution of 3.5 µeV (FWHM). SPHERES is a pure backscattering spectrometer with a dynamic range of +/-31 µeV and a typical resolution of 0.67 µeV (FWHM). To improve statistics near the elastic line, we used the instrument with a reduced dynamic range of +-10.5 ueV The Q-range of both instruments is set by the Si-111 analyser crystals to ~1.8 Å$^{-1}$. Measurements on powder samples have been taken at various temperatures below 100 K.

## III. EXPERIMENTAL RESULTS

### A. Magnetic investigations

All orientation dependent magnetic susceptibilities ($\chi$) of YBaCo$_3$FeO$_7$ were extracted from a single crystalline sample and the data are shown in Fig. 3.

At temperatures above 600 K up to 1000 K, the highest accessible temperature, the inverse susceptibility exhibits a linear decrease indicating paramagnetic like behavior (inset Fig. 3a). Applying the Curie-Weiss law at highest temperatures, *i.e.* between 900 and 1000 K for the low field (50 mT) data, one obtains a Curie-Weiss temperature, $\theta_{CW} = -2.0 \times 10^3$ K and a total effective magnetic moment, $M_{Co,Fe} = 14.5$ $\mu_B$ per formula unit. When taking into account only spin moments, the magnetic moment for Co$^{2+}$ and Fe$^{3+}$ ions is 17.54 $\mu_B$. The caveat is that the estimated values of $M_{Co,Fe}$ and $\theta_{CW}$ may be affected by relatively large systematic errors, because the Curie-Weiss law is expected to be valid in the limit of T>>|$\theta_{CW}$|, which would be above the accessible temperature range.

Near 590 K ($T_{f1}$), there is an obvious transition from isotropic to a weak anisotropic magnetic behavior for the two measured crystallographic directions: on cooling, $\chi$ increases more strongly along the unique axis (*c*) than perpendicular to it. Below $T_{f1}$ field cooled (FC) data and zero field cooled (ZFC) data measured in B = 1 T perfectly superimpose. The small increase in $\chi$ with B$\perp c$ is likely to arise from a slight misalignment of the single crystal, as a resulting magnetic moment perpendicular to *c* is prohibited by the hexagonal symmetry. Magnetic relaxation measurements at 560 K, not shown, gave no indications of a glassy state. We note that for the $^{57}$Fe enriched Mössbauer sample, $\chi$ in a low *DC* field (not shown) also displays the magnetic transition at 590 K, which enables the combination of data from isotope enriched and single crystalline samples.

Considering the low temperature data, shown in Fig. 3b, the relevance of the magnetic history of the sample is obvious, which prevents a quantitative discussion. However, close to 50 K ($T_{f2}$) a second anomaly with a splitting of FC and the ZFC data indicates a further magnetic transition to a glassy-like magnetic state.

The magnetization as function of field was measured at different temperatures both with the applied field parallel and perpendicular to the crystallographic *c*-axis (Fig. 4). For B||*c*, all data below 590 K reveal a weak magnetic hysteresis (Fig. 4a), with an increasing coercive field down to 280 K, a coercive field that, even at lower temperatures, stays almost constant (inset in Fig. 4a). At low temperatures, these hystereses loose their defined shape but have almost the same isothermal magnetic remanence. As the field was scanned at the same rate at all temperature, this behavior corresponds to a ferromagnetic component along *c* below 590 K, which gradually becomes viscous with decreasing temperature. In contrast, the hysteresis is absent if the field is applied perpendicular to the *c*-axis (Fig. 4b), except for a small anomaly most likely resulting from a minor misalignment of the crystal. A direct comparison of data from 560 K for both field directions is shown in inset Fig. 4b to illustrate the anisotropy of the hysteresis. In all applied fields, the induced magnetization is relatively low in agreement with the large negative Curie-Weiss temperature and the related strong antiferromagnetic exchange.

Note that the lack of inversion symmetry in the *Swedenborgite* structure allows for Dzyaloshinsky-Moriya (DM) interactions.[29,30] DM interactions provide a possible explanation for the observed anisotropy and the magnetization hysteresis, where the induced weak ferromagnetic component can correspond to a small spin canting along the structural polar axis.

## B. Mössbauer spectroscopy

The 4.2 K Mössbauer spectrum (Fig.5a) reveals two distinct magnetic sublattices with hyperfine fields of 53 and 43 T. Upon heating, the subspectrum with the smallest field reveals a transition to a paramagnetic state above 50 K and that with the largest field reveals a gradual decrease of the hyperfine field up to 480 K. These observations are consistent with the magnetic susceptibility data that reveal a transition around 50 K and a second transition at 590 K. The spectral areas at 4.2 K indicate that 17(2) % of Fe is located on the site with the larger hyperfine field, which has to be the (Co,Fe)2 site in the kagome plane according to the distribution of Fe in the structure.[18] A Mössbauer spectrum, not shown, recorded at 4.2 K on a thin non iron-57 enriched sample reveals a similar preferential occupation of 19(2) %. This observation indicates an occupation of 82(2)% of Fe on the site (Co,Fe)1 and 6(1)% of Fe on the three sites (Co,Fe)2; as no superstructure reflections are seen in scattering experiments, Fe is assumed to be randomly distributed. However, because of the different valence of $Co^{2+}$ and $Fe^{3+}$,[24] an avoidance for Fe-Fe nearest neighbors is expected in the kagome plane.

The Mössbauer spectra between 4.2 and 480 K were fitted with a model that includes four components: one (Co,Fe)2 sextet, two (Co,Fe)1 sextets below 60 K, two (Co,Fe)1 doublets at and above 60 K, and one small impurity sextet with only 2.5 % spectral area. We cannot identify this impurity explicitly, but tentatively propose $YBa(Co,Fe)_2O_5$ because the hyperfine field and its temperature dependence (*ca.* 55 T at 4.2 and *ca.* 38 T at 480 K) are similar to what is observed in $TbBaFe_2O_5$[31] and are not compatible with a possible $\alpha$-$Fe_2O_3$ impurity retained from the starting material. Moreover, this would agree with the observations from the X-ray diffraction data, as only the enriched sample contains this impurity. The (Co,Fe)1 and (Co,Fe)2 sextets are broadened, which can be attributed to a distribution of the hyperfine fields associated with the binomial distribution of Fe at the two sites. The spectra are not sufficiently resolved to warrant a fit by a binominal distribution; instead, an incremental line-width broadening for the magnetic components was applied. For the (Co,Fe)1 site it was necessary to model the subspectrum with two broadened sextets in the high temperature phase. The reported hyperfine field (Fig.5b) and quadrupole shifts, see below, are the weighted averages.

The isomer shift of the (Co,Fe)2 site in Fig.5c, was fitted with a Debye model, and the associated Lamb-Mössbauer temperature is $\Theta_{LM} = 440(25)$ K. A similar fit of the (Co,Fe)1 site isomer shift, up to 400 K, indicates a significantly lower $\Theta_{LM} = 240(60)$ K, and the variation of the isomer shift above 400 K is anomalous. This large difference in the $\Theta_{LM}$ is surprising, considering that the oxygen coordination of both sites is tetrahedral. The quadrupole shifts for the magnetic subspectra are essentially temperature independent, $QS_{Fe(1)} = -0.06(2)$ mm/s, $QS_{Fe(2)} = -0.11(4)$ mm/s. The average quadrupole splitting of the (Co,Fe)1 site in the paramagnetic phase is $\Delta E_Q = 0.57(2)$ mm/s. The temperature dependences of the hyperfine field for both sites are roughly consistent with a mean-field model, displayed by lines in Fig.5b, for a spin $J = 5/2$ and a critical temperature $T_{f2} = 55$ K and $T_{f1} = 590$ K, see blue and red lines, respectively. Note the acceptable agreement even though these curves are not fitted to the data.

The spectral parameters for the (Co,Fe)1 and (Co,Fe)2 components are typical for high-spin $Fe^{3+}$ in oxygen environment, and compatible with tetrahedral coordination, although the quadrupole splitting for the (Co,Fe)1 site is somewhat smaller than is observed in the paramagnetic spectra of iron garnets.[32] None of the recorded spectra is indicative of localized magnetic fluctuations. Hence, if fluctuations are present, they must be slower than the detection limit of *ca.* 1 MHz for Mössbauer spectroscopy. The observation of a magnetic subspectrum for the (Co,Fe)2 site up to 480 K is consistent with the enhanced magnetic susceptibility observed below 590 K.

Between these two magnetic anomalies (50 – 590 K), the following observations are made: The macroscopic magnetization data, corresponding to large time scales (Hz), exhibit a hysteretic low moment of about 0.01 $\mu_B$ per formula unit along the crystallographic *c*-direction. The Mössbauer spectral data indicate a clear preference of $Co^{2+}$ and $Fe^{3+}$ ions for the kagome and interlayer sites, respectively. Above approximately 590 K, vanishing hyperfine fields suggest that all spins are paramagnetic. Below this anomaly, the observed anisotropy in magnetization relates to the appearance hyperfine fields on $Fe^{3+}$ kagome spins. With respect to the local sensitivity of the technique, we can infer local moments, which are sufficiently static to impose a Zeeman splitting of the nuclear moments. The question whether this behavior is shared by the majority of the $Co^{2+}$ kagome spins cannot be clarified with the present data. The simultaneous onset of anisotropy, of weak magnetization hysteresis, and of hyperfine fields is remarkable, and its origin is not yet clear. Long-range magnetic order as well as any static strong spin correlations at such high temperatures can be ruled out by the neutron scattering results as will be discussed below.

### C. Diffuse polarized neutron scattering

Diffuse polarized neutron diffraction has been applied to study spin correlations and possible magnetic ordering at lower temperatures. The magnetic scattering (see Fig. 6a) has been obtained by polarization analysis. The data were calibrated by the separated isotropic spin-incoherent scattering from the Co ions.

At lower temperatures the most remarkable feature is the wave-vector dependence of the scattering. The observed peaks are broad and diffuse, indicating short-range magnetic correlations rather than long-range magnetic order. With decreasing temperature, the dominant peak near $Q_{c1} = 2\pi/d \approx 1.38$ Å$^{-1}$ evolves continuously and, on further cooling, a second minor peak appears near 50 K, at $Q_{c2} \sim 1.7$ Å$^{-1}$. These most significant changes in the scattering pattern $S(Q)$ occur near the same temperature, at which we observe the appearance of the second hyperfine field in Mössbauer spectroscopy. The peaks are not resolution limited; at 1.2 K, the width of the peak at $Q_1$ fitted by a Lorentzian yields an exponential decay length of 13.5(8) Å (=1/HWHM) for the spin correlations. The correlation length varies only slightly with temperature decreasing smoothly to 10(1) Å at 100 K. This indicates that no true long range magnetic ordering takes place, an observation that is surprising in view of the large (negative) Curie-Weiss temperature, (Fig. 3a). The suppression of long-range order, most likely by the geometrical frustration of the magnetic sites due to the tetrahedral coordination, is certainly one of the strongest observed so far, and can be described by a frustration parameter $f > |\theta_{CW}| / 1.2$ K $\sim$ 1000. Moreover, the comparably low intensity towards small $Q$ reflects that the sum of magnetic moments of neighboring spins is close to zero and in accordance with strong antiferromagnetic exchange. Although $Co^{2+}$ and $Fe^{3+}$ carry different magnetic moments, the intensity at low $Q$ does not show any indications of uncompensated ferri-magnetic moments.

Whereas the Mössbauer hyperfine field at higher temperatures show static (Fe-)moments on the kagome sites, a time-dependence of the pair correlation function of these moments can be inferred from the temperature dependence of the neutron diffraction data indicating significant inelastic scattering contributions on a THz-scale at ambient and higher temperatures.[33]

The details of the low temperature magnetic scattering, in particular the exact peak position $Q_1$ and the more pronounced features in the scattering with the appearance of the second peak at $Q_2$, differs from the diffuse magnetic scattering observed in the *Swedenborgite* $Y_{0.5}Ca_{0.5}BaCo_4O_7$.[17] Therefore, spin-correlations within decoupled two dimensional kagome lattices will not account for the observation; $Q_1$ is close although, significantly different from the value 1.33 Å$^{-1}$ that is expected for the (2/3,2/3,0) ordering wave-vector of the two dimensional √3×√3 ground state of the kagome antiferromagnet.[34] An alternative description will be discussed below.

### D. High-resolution inelastic neutron scattering

The spin dynamics have been further investigated by inelastic neutron backscattering. An overview on the quasi-elastic scattering $S(Q,E)$ *versus* temperature has been obtained using the BASIS spectrometer. Quasi-elastic scattering appears between 70 K and 30 K, see Fig.7a. The $Q$-dependence of this quasi-elastic scattering closely resembles the $S(Q)$ observed in diffraction data measured at DNS. At lower temperatures, all quasi-elastic scattering narrows into the elastic resolution window of 3.5 µeV (FWHM). As there is no indication of structural relaxation, we attribute this to spin freezing.

A complementing study of this apparent freezing transition has been performed at the SPHERES instrument focusing on a narrow dynamic range of +/- 10.5 µeV. Measurements were made upon cooling between 93 K and 4 K at twelve different temperatures. The quasi-elastic intensities have been determined by normalization to monitor counts separating a constant background and integrating over a Q-range from 1.2 to 1.7 Å$^{-1}$. In this Q-range, the change of magnetic scattering intensities with temperature is most significant (Inset (b) Fig.7).

The data measured at 93 K represents essentially the part of the nuclear scattering, mostly spin-incoherent scattering from Co and also the part of possibly already frozen spin correlations. The shape closely resembles the instrumental resolution function, to first order a Gaussian with 0.65 µeV FWHM. Upon cooling the integrated intensity increases and we only note a minor change of the line-width. This is exemplified in a more detailed analysis, presented in Fig.7b, by focusing on a narrow dynamic range of 6 µeV and showing data in the transition regime at 39 K (magnetic and nuclear) and at 93 K (nuclear); the data at 39 K show a relative increase in peak intensity by a factor 1.38 and can be well described by a Gaussian of 0.72 µeV (FWHM). The difference of the intensities taken at 39 K and 93 K is related to only magnetic scattering from spins, which are freezing in, and is in accordance with the increased Mössbauer hyperfine field observed on the Fe located at the (Co,Fe)1 site. The shape of this magnetic intensity can be modeled by a Gaussian of 0.80 µeV (red line in Fig.7b. The slight increase in line-width compared to the instrumental resolution may still indicate slow spin fluctuations. However, no systematic deviations from the Gaussian modeling of the magnetic intensity appears in the difference plot, which is shown in the lower part of Fig. 7b. Furthermore, we cannot achieve any significant better description by modeling a possible exponential time decay of spin correlations using a Voigt-profile. One may note that apart from absorption also extinction due to increased scattering power will affect the observed broadening in resolution, since backscattered neutrons pass twice through the sample. These results are qualitatively representative for the spectra taken at other temperatures.

Both, the BASIS and the SPHERES data reveal a rapid change in the spin dynamics, which indicates a dynamically inhomogeneous behavior. A possible explanation could be that instead of a continuous slowing down upon cooling, sequentially more and more spins freeze in. This is in accordance with the DNS data, which are obtained in a coarser energy integration of a few meV. Here, spin correlations develop at low temperatures, near 100 K, with a significant increase near 50 K, where we also observe anomalies in magnetization with a splitting of field-cooled and zero-field susceptibilities, the onset of Mössbauer hyperfine fields for Fe on the interlayer sites, and the freezing dynamics seen in the high-resolution inelastic neutron scattering. According to these observations, obtained on rather different time scales, we see that when interlayer site moments are getting involved in the spin correlations, this leads to a frozen disordered magnetic structure.

## IV. MONTE CARLO SIMULATIONS

Monte Carlo simulations can provide further insight into characteristic spin structures, especially if the structures are more complex than those of ideal magnetic order, and predictions from simple model Hamiltonians can be compared with the measured scattering. Recently,[16] spin correlations and effects of geometric frustration in the extended kagome lattice of the *Swedenborgite* structure have been studied based on a classical Heisenberg Hamiltonian

$$H = -\frac{1}{2}\sum_{i,j} J_{i,j} \mathbf{S}_i \cdot \mathbf{S}_j \ ,$$

where the exchange $J$ have been confined to nearest neighbor bonds. A distinction of the nearest neighbor exchange for the in-plane coupling between neighboring kagome spins and the out-of plane coupling between kagome and trigonal interlayer sites can be made. Remarkably, as shown by Khalyavin *et al.*,[16] three dimensional long-range ordered ground states cannot form, if the nearest neighbor exchange $J_{in}$ (Co2-Co2, Fig.1b) is larger than the exchange $J_{out}$ (Co1-Co2, Fig.1b). Here, we have varied the ratio $J_{out}/J_{in}$ and calculated the scattering cross-section $S(Q)$ to compare them with the neutron scattering data, especially for low temperatures. The simulations, based on a standard Metropolis algorithm, have been carried out in a super-cell of 12×12×24 unit cells with 8 different magnetic sublattices, and usual periodic boundary conditions have been used; a larger dimension in *c* direction has been chosen because of the rather anisotropic decay of the correlations. Typical configurations at low temperatures were obtained by sequential cooling from high temperatures and by performing more than $10^5$ Monte Carlo steps per spin. For simplicity and in lack of experimental input, Co and Fe were treated as equal magnetic entities. However, when calculating $S(Q)$ we used the site occupancies as determined from the Mössbauer spectroscopy with spin only moments 3/2 and 5/2 for Co and Fe, respectively, and included the magnetic form factors.

A first result is that upon varying the ratio of in-plane and out-of plane coupling strength, for $J_{out}$= 0.5 $J_{in}$ the wave-vector dependence of the scattering can be essentially reproduced for low temperatures. The data shown in Fig.6b correspond to $T = 0.01\ J_{in}/k_B$, a part of the simulated supercell and its spin structure is depicted in the inset. The spin arrangements are not truly long-range ordered in three dimensions. Particularly strong spin correlations along the *c*-direction are found between the trigonal interlayer sites. In the *a-b* plane, at least at shorter distances, one may note a typical 120° rotation between the spin orientations for neighboring trigonal interlayer sites (see Fig. 6b blue spins), which is mediated by neighboring spins on the kagome sites (same figure, red spins). On the other hand, the kagome spins are less ordered along the *c*-direction and their projection does not exhibit any clear pattern. The present findings agree well with the previous studies[15,16] for only nearest neighbor exchange the geometric frustration in the *Swedenborgite* structure promotes strong quasi 1-D spin correlations in *c*-direction. Whereas antiferromagnetically coupled nearest neighbor spins suffer from geometric frustration resulting in a strong damping of spin-correlations in the kagome layers, a ferro-alignment of second neighbors along the polar axis on interlayer sites becomes favorable. However, YBaCo$_4$O$_7$ has a stronger ordering tendency and equal in-plane and out-of plane coupling strength have been proposed based on simulations. In YBaCo$_3$FeO$_7$, the out-of plane exchange $J_{out}$ should be weaker. Our Monte Carlo simulations, using a Heisenberg model with $J_{out} = 0.5\ J_{in}$, show that the observed magnetic scattering and its Q-dependence can be reasonably described and confirm the reported characteristic anisotropy of the spin-correlations in the *Swedenborgite* structure.[16]

## V. DISCUSSION AND CONCLUSIONS

We have presented a comprehensive study of the magnetic properties of $YBaCo_3FeO_7$, a new compound of the *Swedenborgite* structure, of which we have been able to grow large single crystals by floating zone mirror technique. High temperature susceptibility measurements between 600 and 1000 K, yield a Curie-Weiss temperature estimation of ~2000 K related to a strong antiferromagnetic exchange, similarly as been observed previously in the other Co-based *Swedenborgites*, *i.e.* $Y_{0.5}Ca_{0.5}BaCo_4O_7$.[18]

Near 590 K, a magnetic anomaly occurs, which we attribute to a cooperative phenomenon due to the continuous and spontaneous change, clearly deviating from the expected paramagnetic behavior. Using single crystals, we found a weak anisotropic susceptibility with a stronger increase upon cooling along the *c*-axis than in the *a-b* plane and, for temperatures lower than 590 K, a weak magnetization hysteresis for fields B along the *c*-direction. Mössbauer spectroscopy reveals the onset of hyperfine field splitting near to this 590 K anomaly in the susceptibility and magnetization measurements. A quantitative analysis reveals that the hyperfine fields are related to the kagome sites. Furthermore, the trigonal interlayer sites experience a hyperfine field splitting at much lower temperatures, below ~ 50 K. Such a large difference in the temperature at which a hyperfine field appears on two different sites in the same structure is rare. In the $Fe^{2+}Fe_2^{3+}F_8(H_2O)_2$ weak ferromagnet, for example, the $Fe^{3+}$ spins order at 157 K, whereas the $Fe^{2+}$ spins remain idel and order only at ~35 K, *i.e.* a 5 times lower temperature.[35] This idle spin behavior of the Fe(II) spins was seen as an attempt of the system to minimize the frustration by maintaining these spins in a paramagnetic state.[34] In the title compound, all Fe is trivalent, no true long range order appears and the ratio in the temperatures for appearance of local magnetism is even larger, ~10. The idle spin behavior of the interlayer sites between 590 and 50 K might in this respect be also seen as a way to minimize the frustration. In addition, the Co and Fe site occupancies could be determined from the Mössbauer spectral weights. Accordingly, Fe occupies preferentially the trigonal (Fe,Co-1) site to 82(2) %, and the kagome sites (Fe,Co-2) only to 6(1) %. As Mössbauer spectroscopy is a local probe, the appearance of hyperfine fields cannot be taken as an indication of magnetic order. The neutron diffraction experiments indeed reveal that the system does neither order at such high temperature as 590 K nor even orders at very low temperatures.

Although, it seems likely that there is a common origin of the susceptibility anomaly, the weak anisotropic magnetization and the hyperfine fields, the origin is not yet clear. Such susceptibility anomalies have also been observed in $YBaCo_4O_7$ at 313 K and were in that case related to a structural phase transition from trigonal to orthorhombic symmetry. A structural transition is also found in the *Swedenborgite* $YbBaCo_4O_7$.[36] There are, however, no indications of orthorhombicity in $YBaCo_3FeO_7$. Another speculation to pursue in view of the anisotropy is an orbital ordering transition involving the kagome sites, which in combination with symmetry allows for Dzyaloshinskii-Moriya interactions and might explain the observed weak ferromagnetism.

Interestingly, by Mössbauer spectroscopy we can also distinguish a rather different magnetic behavior of the two different sites, the kagome and interlayer sites. The development of hyperfine fields at the interlayer sites below approximately 50 K coincides well with the freezing of spin correlations, at least within the high resolution of neutron back scattering spectroscopy. Neutron diffraction has been applied in combination with polarization analysis for an accurate separation of the magnetic scattering signal. The change of the magnetic scattering intensity versus temperature can be related to the development of hyperfine fields on the interlayer sites. With respect to Mössbauer spectroscopy, the onset of intensity changes in neutron diffraction takes place at higher temperature due to the shorter probe times. The observed peaks of the magnetic scattering remain significantly broadened compared to the instrumental resolution and do not give any evidence for long-range order even down to lowest temperatures of 1.2 K. Both the details of the wave-vector dependence of the magnetic scattering and the obvious participation of interlayer spins, require that the spin correlations extend throughout the three dimensional magnetic sublattice rather than be confined to two dimensions on independent kagome sheets as has been proposed earlier for $Y_{0.5}Ca_{0.5}BaCo_4O_7$.[17] Both systems do not show magnetic long-range order at low temperatures, however, there are significant differences in the scattering pattern indicating also differences in the spin correlations. Whereas in the present study of $YBaCo_3FeO_7$, Fe-Mössbauer spectral results clarify the situation, single crystal neutron diffraction is still needed to prove whether $Y_{0.5}Ca_{0.5}BaCo_4O_7$ is an ideal kagome system as proposed. In this context it is of particular interest to compare the magnetic scattering to calculations based on simple Heisenberg models as presented recently.[16] From these calculations, one may conclude that the ordering in the pure Co based system $YBaCo_4O_7$ is on one hand near to a degeneracy for all zone boundary wave vectors, favoring a disordered ground state, and is on the other hand close to a selection of a single wave vector, the K-point, leading to a 3D ordered phase. The fragile phase stability at least within nearest neighbor interaction Heisenberg models may also explain the inconsistencies in details of the reported magnetic structure,[12,13,15-17] apart from the known sensitivity to the oxygen stoichiometry.[37]

By Monte Carlo simulations, we obtain a reasonable modeling of the magnetic scattering from $YBaCo_3FeO_7$ by using a classical Heisenberg Hamiltonian with antiferromagnetic nearest neighbor interactions, which are half as strong between kagome and interlayer neighbors as compared to the interaction between only kagome sites, $J_{out} = 0.5\ J_{in}$. A characteristic feature of the spin correlations in the model system is the appearance of strong one-dimensional spin correlations parallel to the c-axis and involving the trigonal interlayer sites, a feature which originates from the specific geometric frustration in the *Swedenborgite* structure.[15,16] This picture seems to apply well to $YBaCo_3FeO_7$, at least, at temperatures below ~ 50 K, whereas at higher temperatures above the interlayer spins will fluctuate more rapidly than the kagome spins, which is not expected from Monte Carlo simulations of the used Heisenberg model. We believe that this model reproduces the essential feature of unusual anisotropic spin correlations at low temperatures, however, it has to be considered as a most simple and preliminary model. One may note that according to this model that implies strongly anisotropic spin correlations for the *Swedenborgite* structure, the estimates for an average correlation length obtained from the powder scattering are less meaningful, since the case of partial order and quasi-1 D order is rather difficult to distinguish from a disordered state with isotropic short-range correlations only. The observed change in the spin dynamics also seems to have its origin in the easy formation of quasi-1D ordered domains, whereas a further 3D ordering is blocked by the requested reorientation of the whole domains. Future input from single crystal data will help to establish a more realistic description.

## ACKNOWLEDGEMENTS

We are grateful for comments from Daniel Khomskii and Gary J. Long. Moulay Sougrati is acknowledged for assisting in the collection of the Mössbauer spectral data. We would like to thank Anne Möchel for help in measuring magnetic susceptibility of the isotope-enriched sample. This work was supported by DFG through the project SFB 608, by the FNRS through Grants No.

9.456595 and 1.5.064.5, and by the Helmholtz Association of German Research Centers through grant NG-407. Part of this research at the instrument BASIS Oak Ridge National Laboratories Spallation Neutron Source was sponsored by the Scientific User Facilities Division, Office of Basic Energy Sciences, U. S. Department of Energy.

Figure Captions:

Fig. 1. (Color online) (a) Polyhedral representation of the hexagonal *Swedenborgite* ($P6_3mc$) YBaCo$_3$FeO$_7$ structure. The oxygen (black) at the corner of the Co,Fe containing tetrahedra are also labeled O1-O3, representing three crystallographically different sites. The coordinations of Ba and Y are indicated with black lines and the unit cell in green. (b) The same structure as in (a) but without oxygen atoms and the *kagome* sublattice is highlighted with dark bonds (red); the brighter bonds (blue) mark the interlayer couplings. The Wyckoff labels for the transition metal sites are stated within parentheses. A Y,Ba substructure has been created with thin bonds to emphasize the similarity between the cationic structure in *Swedenborgites* and the close packed Laves phases.

Fig. 2. (Color online) X-ray powder diffraction from a ground part of the single crystal (red lines) and the iron-57 enriched sample (blue lines). The inset shows small peaks from two different impurities, designated **a** (Y$_2$O$_3$) and **b** (YBa(Co,Fe)$_2$O$_5$). A photo of the single crystal is added as inset and a finger is present for size comparison.

Fig. 3. (Color online) Temperature dependent inverse magnetic susceptibility ($\chi^{-1}$) of YBaCo$_3$FeO$_7$ at high (a) and low (b) temperatures. The curves are indicated with B||$c$ if the magnetic field was applied parallel to the $c$-axis and B$\perp c$ perpendicular. Both field cooled (FC) and zero field cooled (ZFC) data are presented at all temperatures. The inset shows susceptibility data at a smaller field of 50 mT extending to 1000 K. The dashed line represents a Curie-Weiss extrapolation from the data between 900 and 1000 K.

Fig. 4. (Color online) Field dependent magnetization in $\mu_B$ per formula unit (FU) of YBaCo$_3$FeO$_7$ at several temperatures and with (a) the applied field parallel (B||$c$) and (b) perpendicular (B$\perp c$) to the $c$-axis. The inset in (a) exhibit measurements at lower temperatures and the inset in (b) display the different behaviors depending on field direction at 560 K.

Fig. 5. (Color online) (a) Mössbauer spectra from YBaCo$_3$$^{57}$FeO$_7$ at selected temperatures. Spectra in the left panel are vertically enlarged in order to better display the magnetic subspectra. The temperature dependence of the hyperfine field (b) and the isomer shift (c) for both sublattices are shown in blue (Co,Fe)$_1$ and red (Co,Fe)$_2$.

Fig. 6. (Color online) (a) Temperature and wave-vector dependence of the diffuse magnetic neutron scattering from a YBaCo$_3$FeO$_7$ powder sample. The separation of magnetic scattering is obtained by polarization analysis at the DNS instrument. A diffuse peak near $Q$ =1.37 Å$^{-1}$ and second peak near $Q$ =1.73 Å$^{-1}$ evolves upon cooling. The finite width indicates short-range correlations rather than 3D ordering at low T. The intensity increase is most likely due to the onset of strong spin correlations perpendicular to the kagome layers (see text). (b) Monte Carlo simulations based on a Heisenberg model, where $J_{out}$ = -0.5 $J_{in}$ for low temperatures, $T = 0.01 J_{in}$. The inset shows a projection along $c$ of a small part of the calculated spin arrangement. The two red colors denote the spins in the alternating *kagome* layers, the blue colors the spins at trigonal interlayer sites. Spins on the trigonal sites (blue) show 1D ordering along the $c$-axis. In the $ab$-plane, a local ordering with ~120° angles can be seen between the (blue) spins on trigonal sites as well as between the mediating kagome spins, which is emphasized by the (black) spins of the top kagome layer.

Fig. 7. (Color online) (a) Intensity maps of the scattering $S(Q,E)$ as measured with the backscattering instrument BASIS. For temperatures between 30 K and 70 K, a change of the magnetic quasi-elastic scattering is observed in the accepted energy range of the instrument with a freezing into the elastic resolution window (~4µeV) at low temperature, 10 K. The intensity peaks at the same $Q$ as observed in the diffraction data taken at DNS. (Color scheme is given in log-scale.) (b) Inelastic scattering on SPHERES indicates spin freezing as exemplified by spectra obtained at 95 K and 39 K for $Q$ from 1.2 Å$^{-1}$ to 1.7 Å$^{-1}$. The difference of these intensities is due to magnetic scattering. The lower part shows the related error bars and the difference to a modeling by a Gaussian on a linear scale (red line, 0.80 µeV FWHM). The inset shows the temperature dependence of the integrated quasi-elastic peak intensity for all measured spectra.

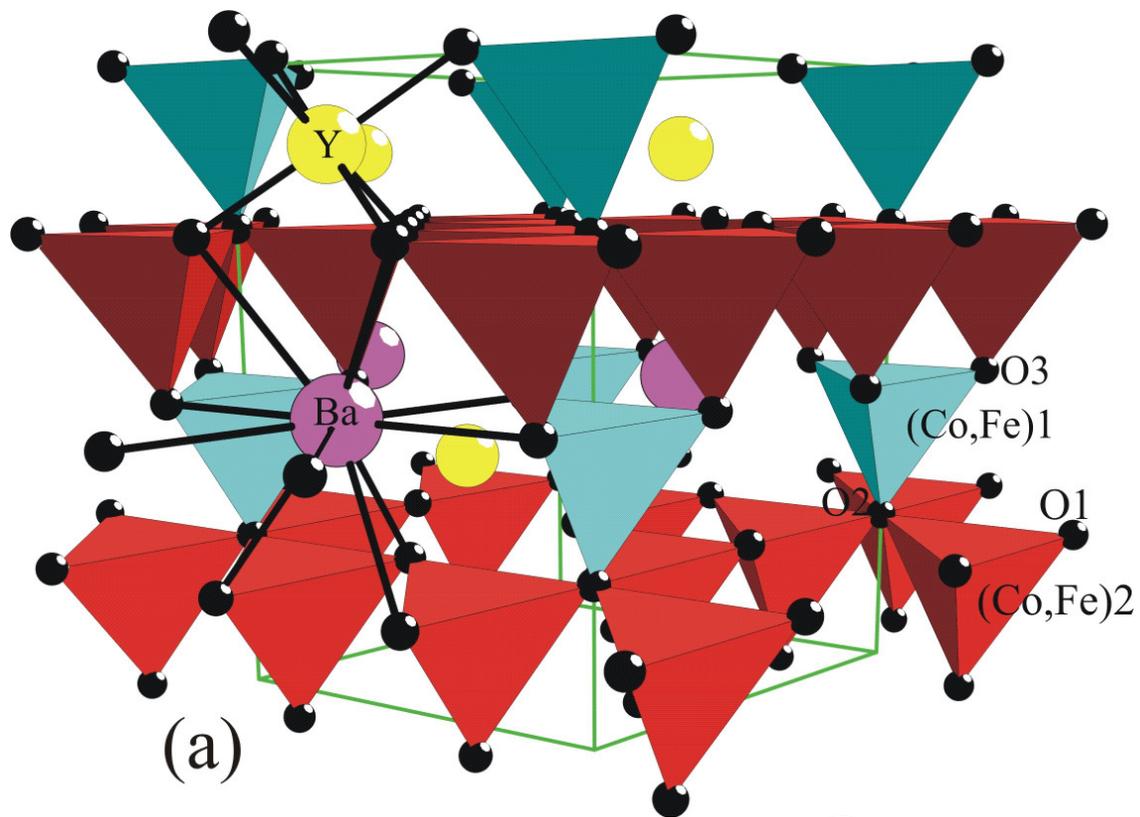

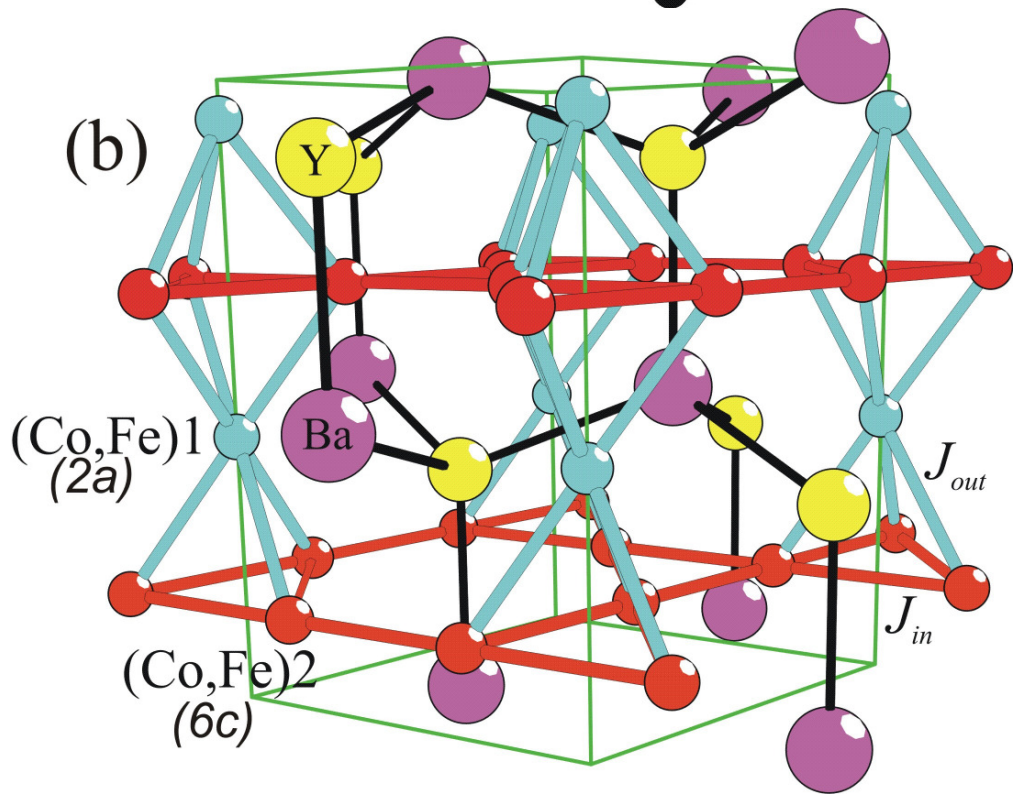

Fig.1

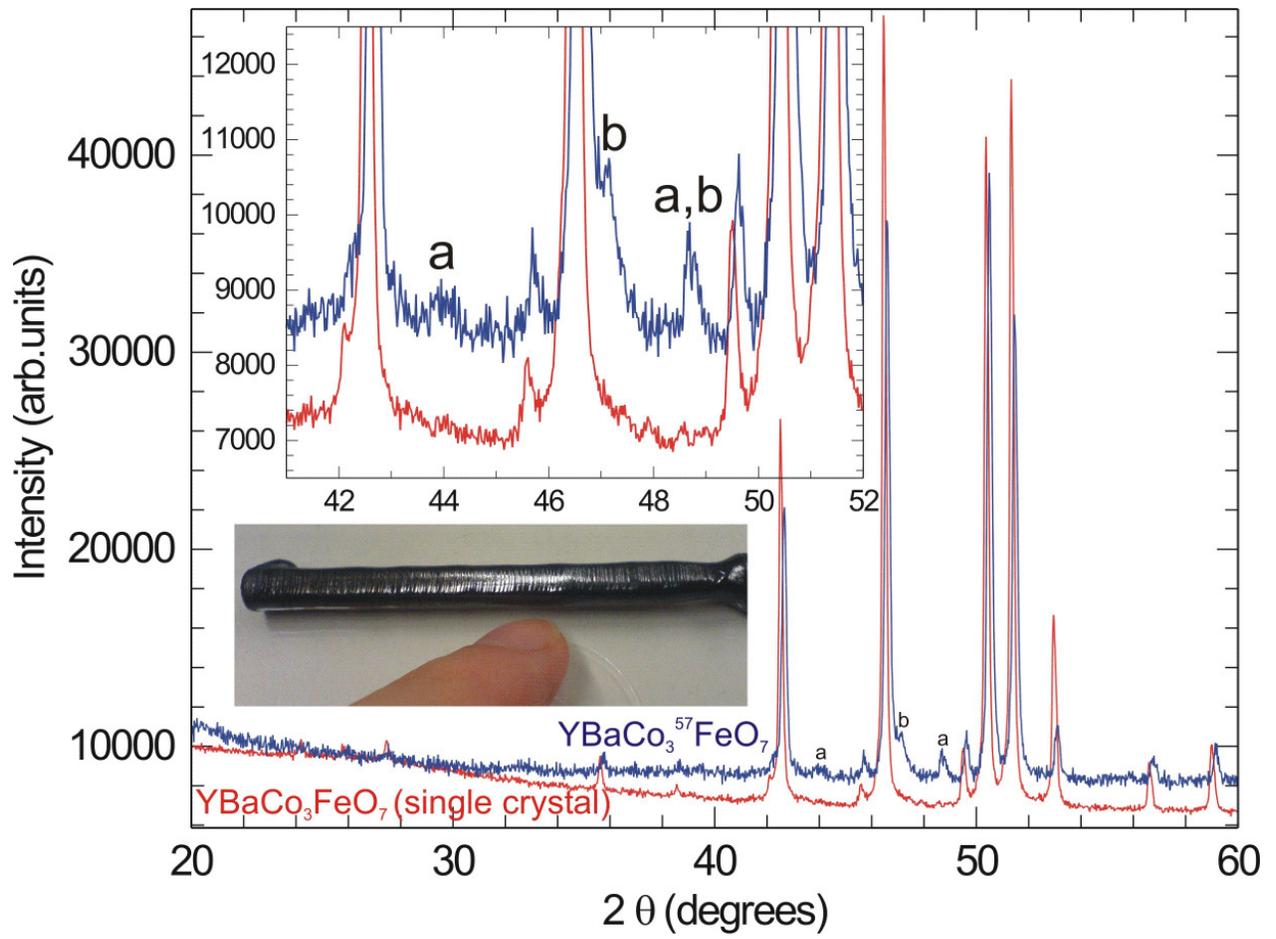

Fig.2

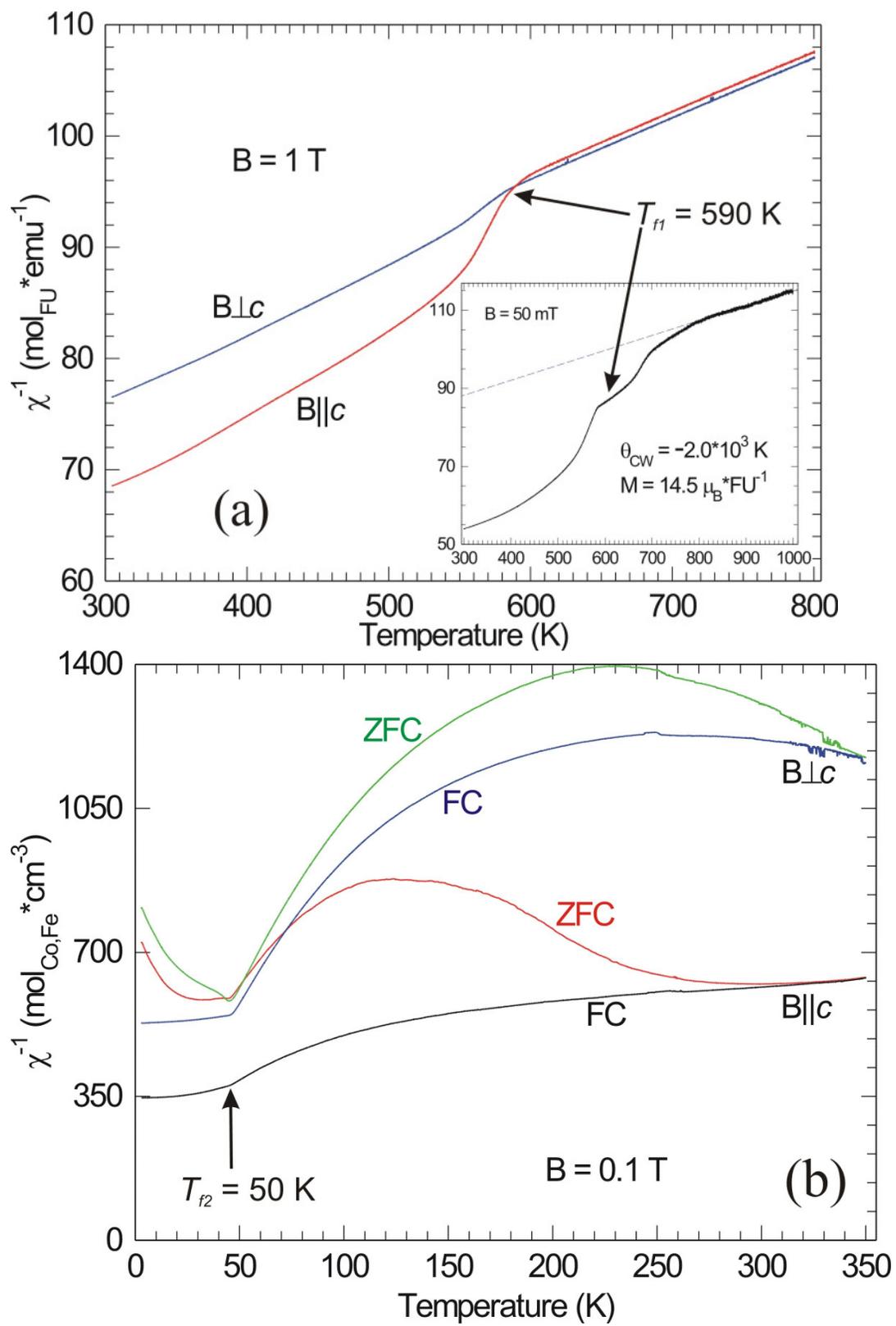

Fig.3

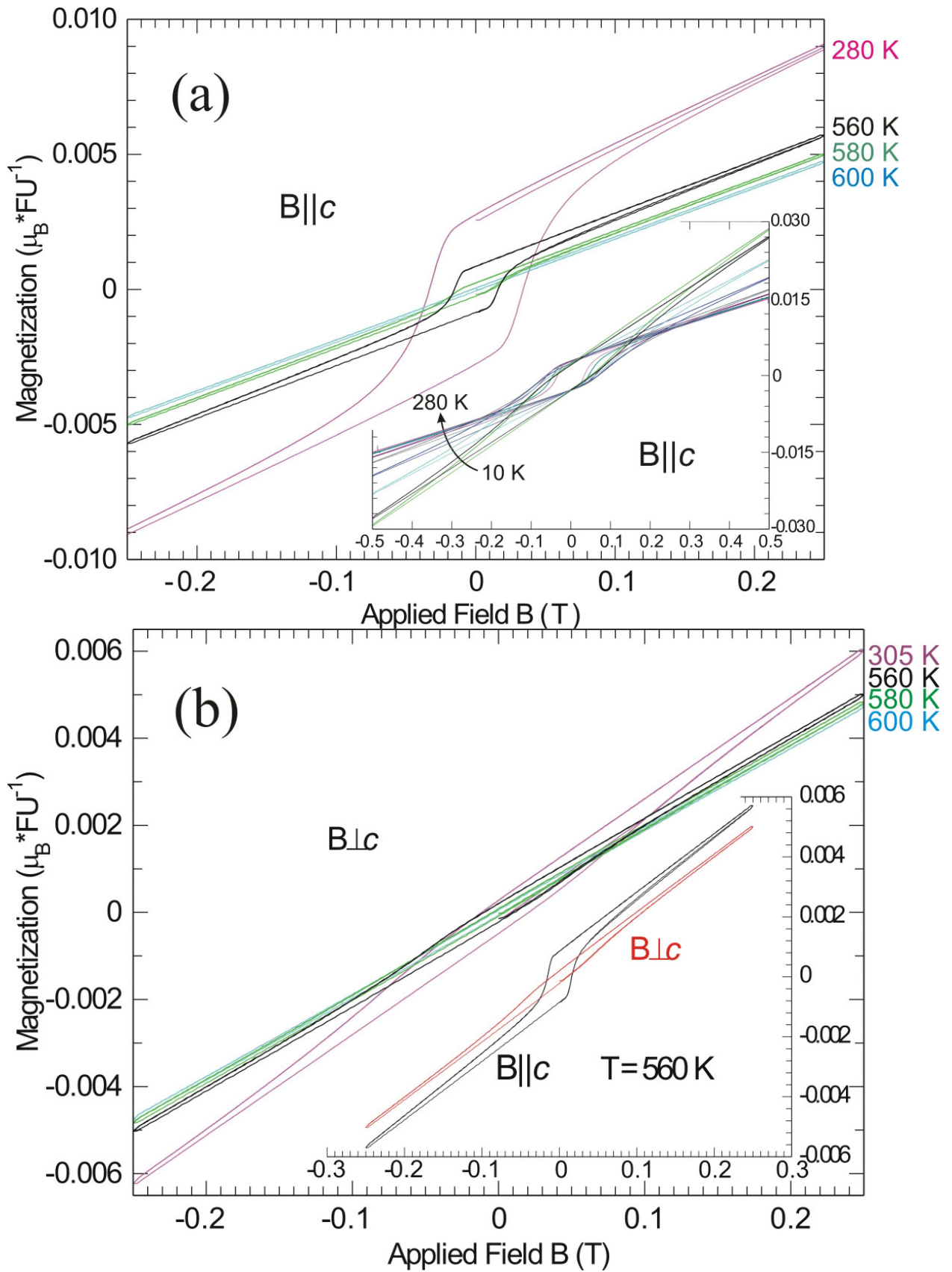

Fig.4

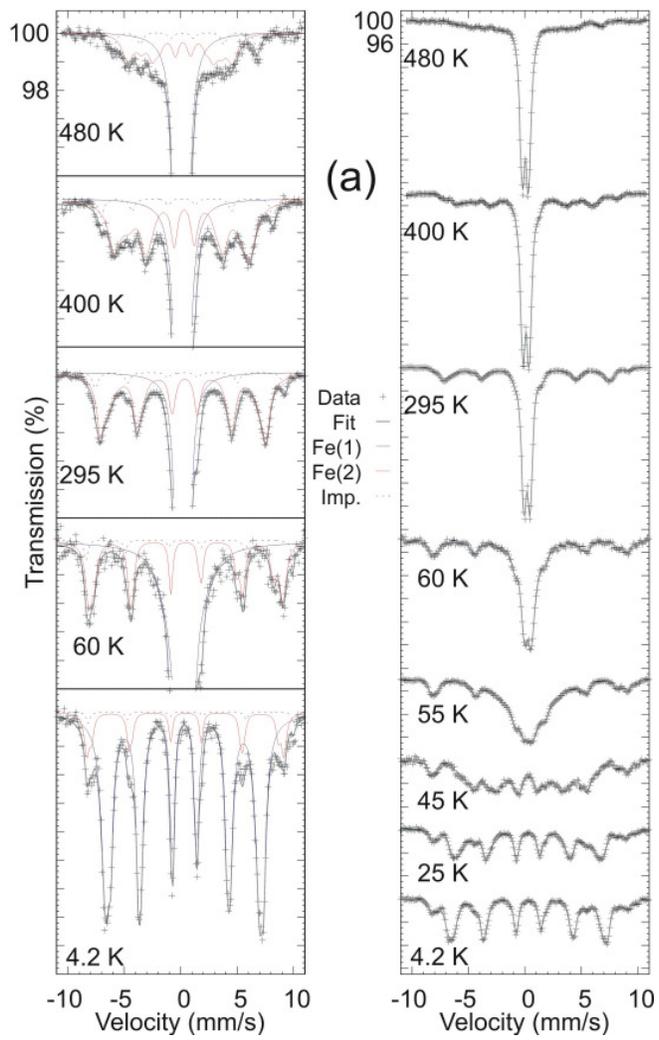

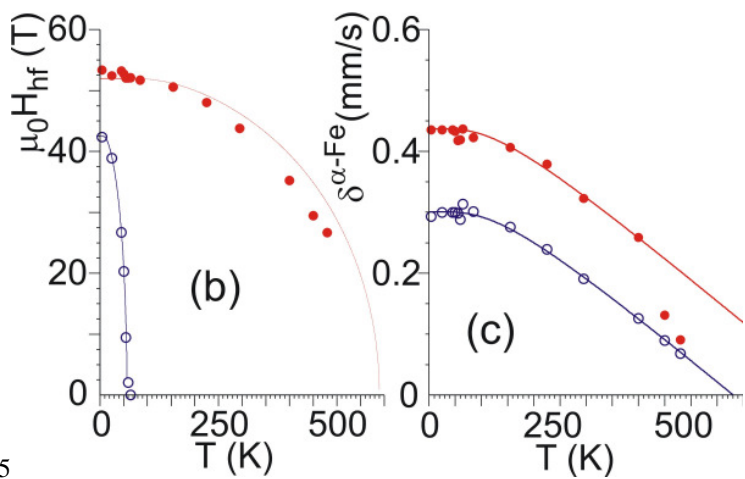

Fig.5

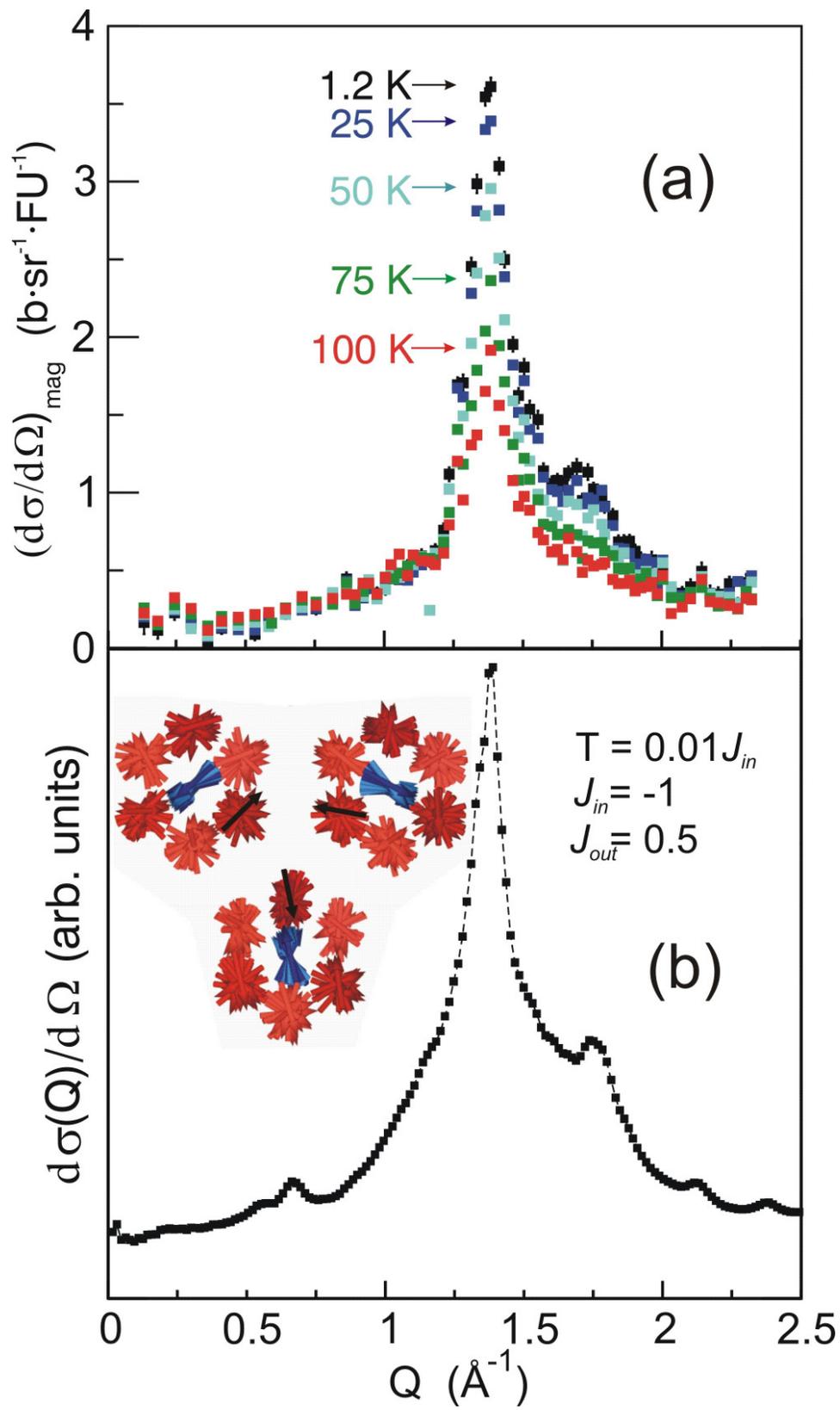

Fig.6

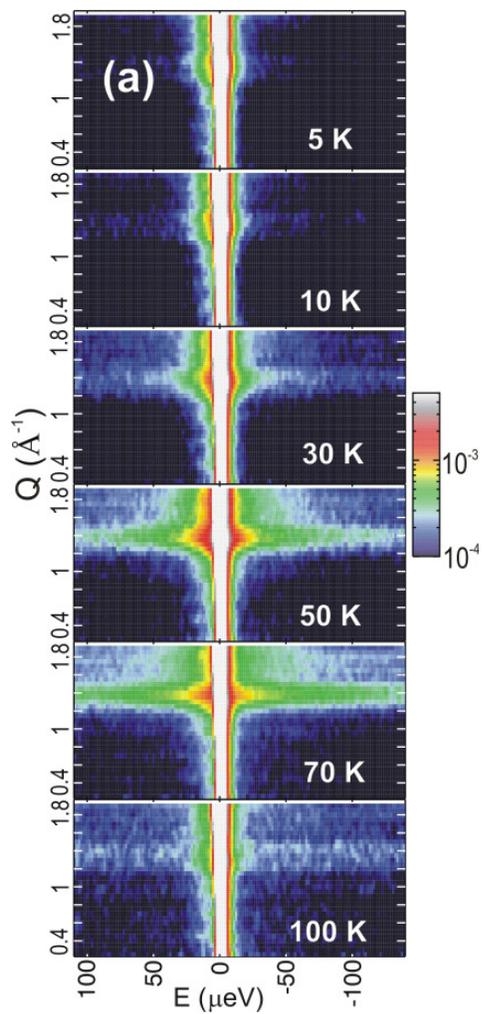

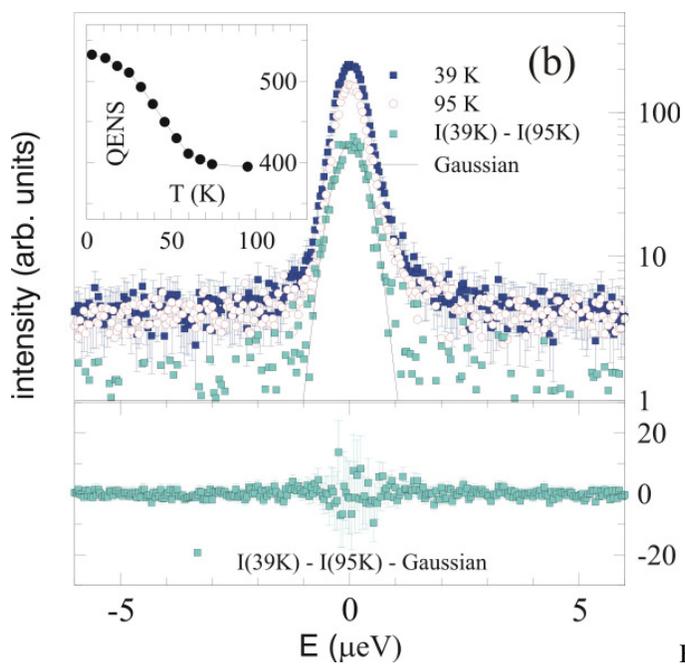

Fig.7